\documentclass[aps,pra,twocolumn,showpacs,superscriptaddress,10pt]{revtex4-2}
\usepackage[colorlinks ,linkcolor=blue,anchorcolor=blue,citecolor=blue,urlcolor=blue]{hyperref}
\usepackage{amsmath}
\usepackage{bm}
\usepackage{graphicx}
\usepackage{epstopdf}
\usepackage{amssymb}
\usepackage{tikz}
\usetikzlibrary{backgrounds}
\usepackage{color}

\begin{document}


\title{Fermionic Lattice Supersolidity in the Attractive Three-Color Fermi--Hubbard Model}

\author{Xiang Li}
\affiliation{School of Physics and Technology, Wuhan University, Wuhan
	430072, China}


\author{Yu Wang}
\email{yu.wang@whu.edu.cn}
\affiliation{School of Physics and Technology, Wuhan University, Wuhan 430072, China}

\begin{abstract}
The recent experimental realization of the half-filled three-color Fermi–Hubbard model on a square optical lattice provides a novel platform for exploring exotic states of matter beyond conventional SU(2) systems. In this Letter, we investigate the three-color Fermi–Hubbard model with color-dependent attractive interactions using determinant quantum Monte Carlo simulations. We find that, at quantum degenerate temperatures, a lattice Fermi supersolid state emerges from the interplay between a moderately-to-strongly interacting two-color subsystem and a weakly coupled third-color environment. This supersolid state, characterized by the coexistence of charge-density-wave and color-superfluid orders, is highly promising for experimental detection with current techniques. Our results demonstrate that the attractive three-color Fermi–Hubbard model on a square optical lattice offers an experimentally accessible system for exploring the supersolidity of ultracold lattice fermions, requiring only a simple lattice geometry and easily tunable onsite interactions.
\end{abstract}

\maketitle
\textit{Introduction—}Supersolidity is a counterintuitive state of matter characterized by the coexistence of superfluid and crystalline orders~\cite{SS1,SS2,Chester1970,SS3,SS4}, which exhibit inherently contradictory behaviors. The supersolid state was first predicted to exist in solid helium over fifty years ago, but definitive experimental evidence for supersolid helium remains elusive. In the past two decades, ultracold atoms and molecules have opened up a new avenue for realizing supersolids. A notable milestone is the experimental realization of supersolids in continuous space (i.e., density-modulated superfluids~\cite{SS2}) in various platforms, including elongated-trap systems~\cite{elongateChomaz2019,elongateTanzi2019,elongateBottcher2019,elongatepollet2019,elongateNorcia2021}, cavity-mediated systems~\cite{cavityLeonard2017,cavityLandig2016,cavityKlinder2015}, spin-orbit-coupled Bose–Einstein condensates~\cite{SOCLi2017}, and the two-dimensional dipolar $^{166}\mathrm{Er}$ gas~\cite{2DStripeSS}. While bosonic supersolidity in continuous space is now well established, the search for supersolid states in ultracold atoms and molecules loaded into optical lattices remains challenging. 

Theoretical proposals for lattice supersolidity have heavily focused on ultracold bosonic atoms in optical lattices, suggesting two possible routes to the relaization of supersolid states: the long-standing hard-core Bose-Hubbard model with additional nearest-neighbor repulsions on a triangular optical lattice ~\cite{hcbBoninsegni2003,hcbMelko2005,hcbHeidarian2005,hcbWessel2005,hcbBoninsegni2005,hcbMurthy1997}, and the recently established two-dimensional Rydberg atom arrays with a $1/r^{3}$-decaying hopping integral and Van der Waals interactions ~\cite{SSRydbergLi2018,SSRydbergPlloet2025,SSRydbergChen2025}. Both of these schemes require off-site interactions to exhibit supersolidity. While off-site interactions occur naturally in Rydberg-atom experiments ~\cite{SSRydbergPeter2012,SSRydbergEbadi2021,SSRydbergweckesser2025}, implementing them with ground-state atoms in optical lattices remains a major challenge. Furthermore, experimentally measuring superfluidity in Rydberg atoms still poses significant technical difficulties. It is known that the extended hard-core Bose-Hubbard model can be mapped to a spin-$1/2$ XXZ model. Excitingly, there exist XXZ-type materials in condensed matter systems, in which spin supersolidity has been discovered experimentally ~\cite{SSCmTsurkan2017,SSCmXiang2024,SSCmChen2026,SSCmZhu2024,SSCmQu2026}.

In this Letter, we explore an alternative route to realizing lattice supersolidity based on the three-color Fermi-Hubbard model (here ``color'' denotes a hyperfine spin component), which has recently been realized in a cubic optical lattice with ultracold $^{173}$Yb atoms at $1/3$ filling~\cite{3compTusi2022} and on a square optical lattice with ultracold $^{6}$Li atoms at half filling~\cite{3compMongkolkiattichai2025}. The three-color model with attractive interactions is theoretically predicted to exhibit an inherent competition between two distinct types of bound states: two-body doublons and three-body trions, which can respectively give rise to a color superfluid (CSF) and a charge density wave (CDW)~\cite{Okanami2014,Miyatake2011,Kantian2009,Capponi2009,3compXu2023,3compLi2023,3compLi2024,3compLi2025,3compStepp2025}. The interplay between the trionic CDW order and itinerant superfluidity raises the possibility of realizing a Fermi supersolid state. To illustrate this, we focus on the half-filled three-color Hubbard model with attractive color-dependent interactions on a square lattice — a model experimentally accessible in Ref.~\cite{3compMongkolkiattichai2025}. Using determinant quantum Monte Carlo (DQMC) simulations, we show that a Fermi supersolid state emerges when the subsystem of moderately-to-strongly interacting color-1 and color-2 fermions is weakly coupled to color-3 fermions — an interaction regime in which three-body loss is significantly suppressed~\cite{3compMongkolkiattichai2025,LiOttenstein2008,LiHuckans2009,LiSchunck2005}. This route to Fermi supersolidity is thus completely distinct from previous proposals based on hard-core bosons and Rydberg atoms, as it requires only a simple lattice geometry and easily tunable on-site interactions, thereby providing a clear pathway for observing Fermi supersolidity in upcoming three-color cold-atom experiments.

\label{sec:model}
\textit{Model—} At half filling, the Hamiltonian for the attractive three-color Hubbard model is defined as: 
\begin{equation}
	H = -t\sum_{\left \langle i,j \right \rangle, \alpha}(c_{i\alpha}^{\dagger}c_{j\alpha} + \mathrm{h.c.})+\sum_{i,\alpha<\beta} U_{\alpha\beta} (n_{i\alpha} - \frac{1}{2}) (n_{i\beta} - \frac{1}{2}).
	\label{eq:hubbard}
\end{equation}
Here, $ \left \langle i,j \right \rangle $ denotes a pair of nearest-neighbor sites on a square lattice, and $\alpha, \beta =1, 2, 3$ represent the color indices. The operators $ c^\dagger_{i\alpha} $ and $ c_{i\alpha} $ correspond to the fermionic creation and annihilation operators, respectively, for a color-$ \alpha $ hyperfine state on site $i$. The particle number operator is defined as $n_{i\alpha}=c^{\dagger}_{i\alpha}c_{i\alpha}$. The hopping amplitude $t$ is set as the unit of energy for the system. The color-dependent Hubbard attractions are denoted by $U_{\alpha \beta} < 0 $, where we set $U_{12}=U$ and $U_{13}=U_{23}=U^{\prime}$.

In the limit $|U'| \to 0$, colors 1 and 2 form an isolated two-component attractive Hubbard model, while color 3 acts as an uncoupled Fermi gas. At half filling, the SU(2) pseudospin-degenerate SF and CDW states of this model are mutually exclusive. When a weak attraction to clor 3 ($|U'|\ll |U|$), the background $12$-CDW generates a slight spatial imbalance in the color-3 density, which in turn deepens the sublattice potential well for $12$-pairs, thereby strengthening the $12$-CDW order. On the other hand, the weak attraction to color 3 acts as a localized potential well that pins $12$-pairs to specific lattice sites, weakening the $12$-SF order. This weak coupling to color 3 may thus drive a transition from mutually exclusive degenerate states into a coexisting supersolid state. 

To demonstrate this route to supersolidity, we employ the determinant quantum Monte Carlo (DQMC) method to simulate the attractive three-color Hubbard model at half-filling. This model is free of the sign problem when the Hubbard-Stratonovich decomposition is performed in the spin-flip channel~\cite{3compXu2023,3compLi2023,3compLi2024,3compLi2025,algorithmWang2015}. The DQMC simulations are conducted on square lattices with linear sizes $L=6,8,10$, and $12$ at the temperature of $T=1/12$, placing the system within the quantum degenerate regime. 

\label{phasestructure}

\begin{figure}[t]
\includegraphics[width=1.0\linewidth]{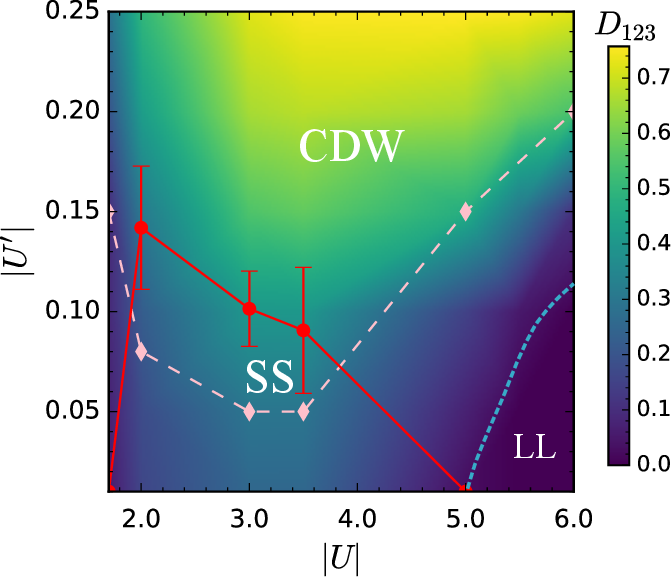}
\caption{Phase diagram of the half-filled three-color Hubbard model on a square lattice at the temperature of $T=1/12$. The color scale indicates the magnitude of the CDW order parameter $D_{123}$. The supersolid (SS) phase is enclosed by the red data points with error bars. The blue dotted curve represents the $D_{123} = 0.03$ contour, inside which a liquid-like (LL) phase exists. The pink dotted curve marks the crossover from $12$-CDW to $123$-CDW.
}
\label{fig:phasestructure}
\end{figure}

\textit{Three-color Fermi Supersolidity—} For $|U'|\ll |U|$, a weak coupling to color 3 can provide just enough density modulation to pin the crystalline order without completely suppressing the superfluidity of color-1 and color-2 fermions. This delicate balance is precisely where the supersolid phase emerges in the phase diagram. To identify this phase, we calculate the charge density wave (CDW) order parameter, which quantifies the density modulation, and the 12-superfluid (SF) correlation ratio, which serves as a reliable marker for the 12-SF phase.  
The CDW order parameter is defined as
\begin{equation}
\label{CDW123}
D_{123}=\sqrt{\frac{1}{N}S_{123}(L,\vec{Q})},
\end{equation}
where $N = L^2$ denotes the total number of lattice sites, and the CDW structure factor, $S_{123}(L, \vec{Q})$, is given by
$S_{123}(L,\vec{Q})=\frac{1}{N}\sum_{ij}e^{i\vec{Q}\cdot (\vec{r}_{i}-\vec{r}_{j})}\sum_{\alpha\beta}\langle n_{i\alpha}n_{j\beta}\rangle.$
Here, $\vec{Q}=(\pi,\pi)$ represents the nesting vector of the square lattice. 
By performing finite-size extrapolations of $D_{123}$ (see Supplemental Material~\cite{SupplementalMaterial} for details), we extract the strength of the CDW order in the thermodynamic limit across various values of $|U|$ and $|U'|$. In Fig.~\ref{fig:phasestructure}, we construct the color map of $D_{123}$ by interpolating discrete DQMC data points across the $(|U|, |U'|)$ plane, providing a smooth visualization of the CDW order strength. The color scale directly corresponds to the magnitude of the CDW order $D_{123}$. Overall, the CDW order is monotonically enhanced by increasing $|U'|$. The region enclosed by the $D_{123} = 0.03$ contour (the blue dotted line) hosts a liquid-like phase. This disordered phase is characterized by a mixture of color-3 fermions, 12-pairs and a small fraction of trions in the weak-$U'$ regime (see the Supplemental Material~\cite{SupplementalMaterial} for details).

The dimensionless CSF correlation ratio is defined as
\begin{equation}
\label{CSForder}
R_{\mathrm{CSF}} = 1 - \frac{S_{\mathrm{CSF}}(L,\Gamma+\delta k)}{S_{\mathrm{CSF}}(L,\Gamma)},
\end{equation}
where $S_{\mathrm{CSF}}(L,\vec{Q})=\frac{1}{N}\sum_{ij}e^{i\vec{Q}\cdot (\vec{r}_{i}-\vec{r}_{j})}\langle c_{i1}^{\dagger}c_{i2}^{\dagger}c_{j2}c_{j1}+\mathrm{h.c.}\rangle $ is the CSF structure factor. 
$\Gamma=0$ is the zone center, and $\Gamma+\delta k$ represents the smallest available neighboring wave vector in the reciprocal lattice. The correlation ratio approaches unity ($R_{\mathrm{CSF}}\rightarrow 1$) when the CSF order is well developed, and vanishes ($R_{\mathrm{CSF}}\rightarrow 0$) in the non-CSF phase. 
The crossing of the $R_{\mathrm{CSF}}$ curves for different system sizes $L$ determines the critical coupling strength $|U'_c|$ (for a given $|U|$) at which the CSF order vanishes (see Supplemental Material~\cite{SupplementalMaterial} for details).
By performing this analysis for various values of $|U|$, we obtain the corresponding critical points $|U_c'|$ (indicated by the red dots in Fig.~\ref{fig:phasestructure}), which delineate the phase boundary between the CSF and non-CSF regimes. 
Because the region enclosed by this boundary exhibits an appreciable $D_{123}$ (Fig.~\ref{fig:phasestructure}), it can be identified as the supersolid phase characterized by the coexisting CSF and CDW orders. 

As $|U'|$ increases, each site hosting a $12$-pair tends to trap a color-3 fermion, forming a localized trion. This drives a crossover from $12$-CDW to $123$-CDW (trion CDW), marked by the pink dotted curve in the phase diagram (Fig.~\ref{fig:phasestructure}), which reveals a broad trion-CDW region within the supersolid phase. The crossover criterion is discussed in Supplemental Material~\cite{SupplementalMaterial}.

\textit{Competition between CSF and CDW orders—} In the supersolid phase, the CSF order parameter is defined as
\begin{equation}
\label{CSForder}
P_{\mathrm{CSF}}=\sqrt{\frac{1}{N}S_{\mathrm{CSF}}(L,\Gamma)}
\end{equation}
Extrapolating $P_{\mathrm{CSF}}$ to the thermodynamic limit yields the CSF order strength across a range of 
$|U|$ and $|U'|$ values (see Supplemental Material~\cite{SupplementalMaterial} for details). 

To systematically examine the competition between CSF and CDW orders, Figs.~\ref{fig:horizontal} (a)-(c) plot $P_{\mathrm{CSF}}$ and $D_{123}$ as a function of interaction strength $|U|$ for fixed couplings $|U'|= 0.05$, $0.10$, and $0.15$. Across all three cases, both $P_{\mathrm{CSF}}$ and $D_{123}$ display a non-monotonic dependence on $|U|$, initially growing in the weak-to-intermediate coupling regime before declining at strong coupling due to thermal fluctuations. Despite this large-$|U|$ suppression, pronounced CSF and CDW orders coexist over a broad range of $|U|$ for weak $U'$, providing clear evidence for a robust supersolid phase. Comparing Figs.~\ref{fig:horizontal} (a)-(c) reveals that increasing $|U'|$ strengthens the CDW order ($D_{123}$) while suppressing the CSF order ($P_{\mathrm{CSF}}$). This competitive mechanism is directly tracked in Fig.~\ref{fig:horizontal} (d) at $|U|= 2$, where growing $|U'|$ drives a monotonic increase in CDW order and a continuous reduction in CSF order, establishing the stability bounds of the coexisting supersolid state.

The suppression of CSF and CDW orders at large $|U|$ can be understood via a low-energy effective Hamiltonian:
\begin{figure}[t]
\includegraphics[width=0.8\linewidth]{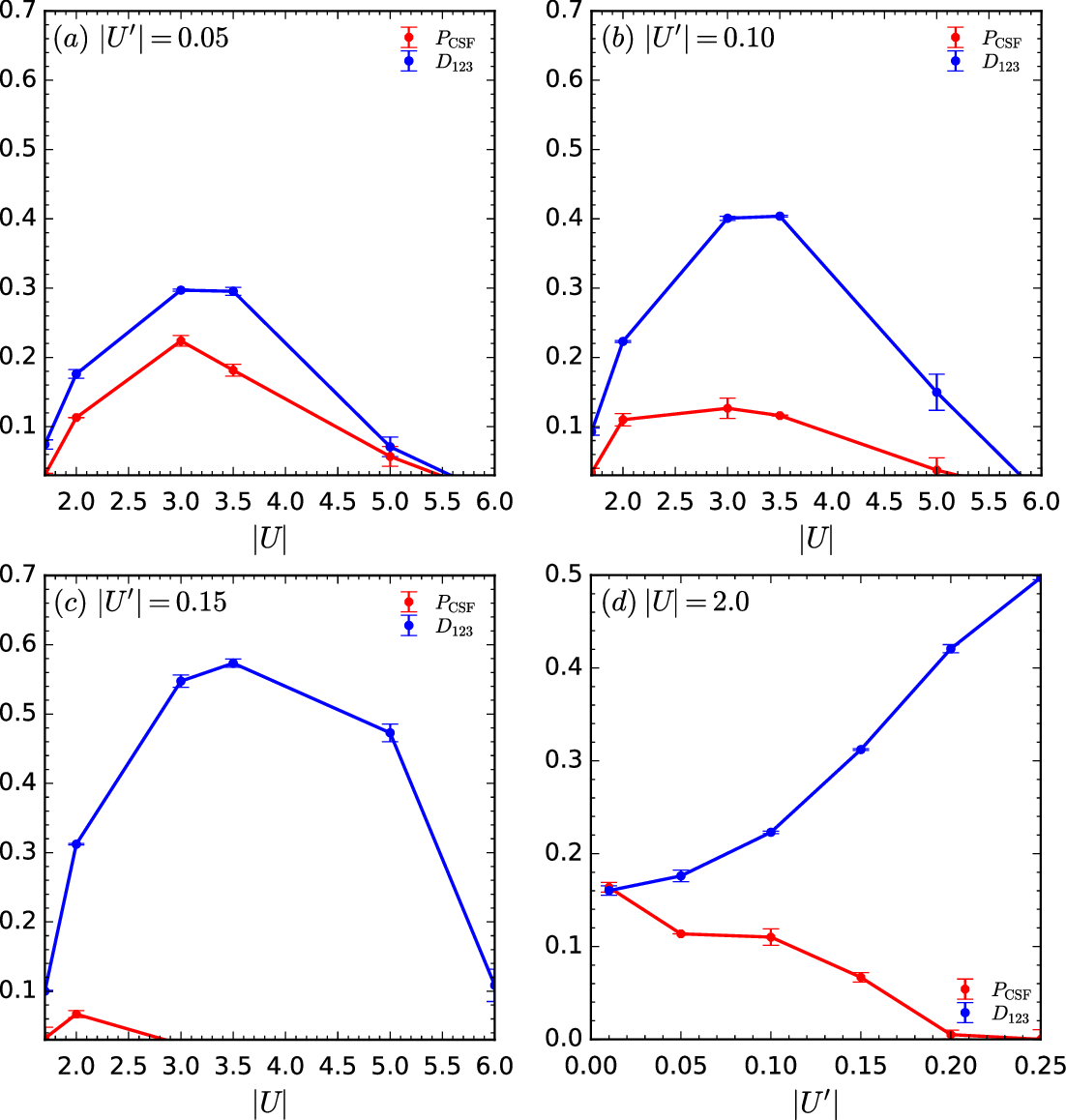}
\caption{Interplay between CDW and CSF orders at half-filling. (a)–(c) CSF ($P_{\mathrm{CSF}}$) and CDW ($D_{123}$) order parameters as a function of interaction strength $|U|$ for couplings (a) $|U'| = 0.05$, (b) $|U'| = 0.10$, and (c) $|U'| = 0.15$. (d) Dependence of $P_{\mathrm{CSF}}$ and $D_{123}$ on $|U'|$ at fixed $|U| = 2.0$. Here, $U$ is the coupling between colors 1 and 2, and $U'$ is the coupling between color 3 and colors 1 \& 2. The temperature is $T = 1 / 12$. 
}
\label{fig:horizontal}
\end{figure}
\begin{equation}
\label{perturbation}
H_{\mathrm{eff}}=-J\sum_{\langle i,j\rangle}(\Delta_i^\dagger\Delta_j+\mathrm{h.c.})+ V\sum_{\langle i,j\rangle}n_in_j,
\end{equation} 
where $\Delta_i^\dagger = c^\dagger_{i1} c^\dagger_{i2}$ ($\Delta_i = c_{i2} c_{i1}$) creates (annihilates) a 12-pair at site $i$, $n_i$ is the local particle number operator, and $J \sim t^2/\vert{}U\vert{}$ and $V \sim t^2/\vert{}U\vert{}$ represent the effective pair-hopping amplitude and nearest-neighbor density repulsion, respectively. At large $\vert{}U\vert{}$, the phase stiffnesses of the CSF and CDW orders are governed by $J$ and $V$, respectively. Because both energy scales vary inversely with $\vert{}U\vert{}$, the orders become increasingly susceptible to thermal fluctuations. This accounts for the strong suppression of the CSF and CDW orders at large $\vert{}U\vert{}$, even deep in the quantum degenerate regime ($T = 1/12$). Additionally, a $12$-pair undergoes decohering scattering from a color-3 fermion~\cite{3compLi2024}, causing the CSF order to decay faster than the CDW order, as shown in Figs.~\ref{fig:horizontal}(a)–(c).   

\begin{figure}[b]
\includegraphics[width=0.7\linewidth]{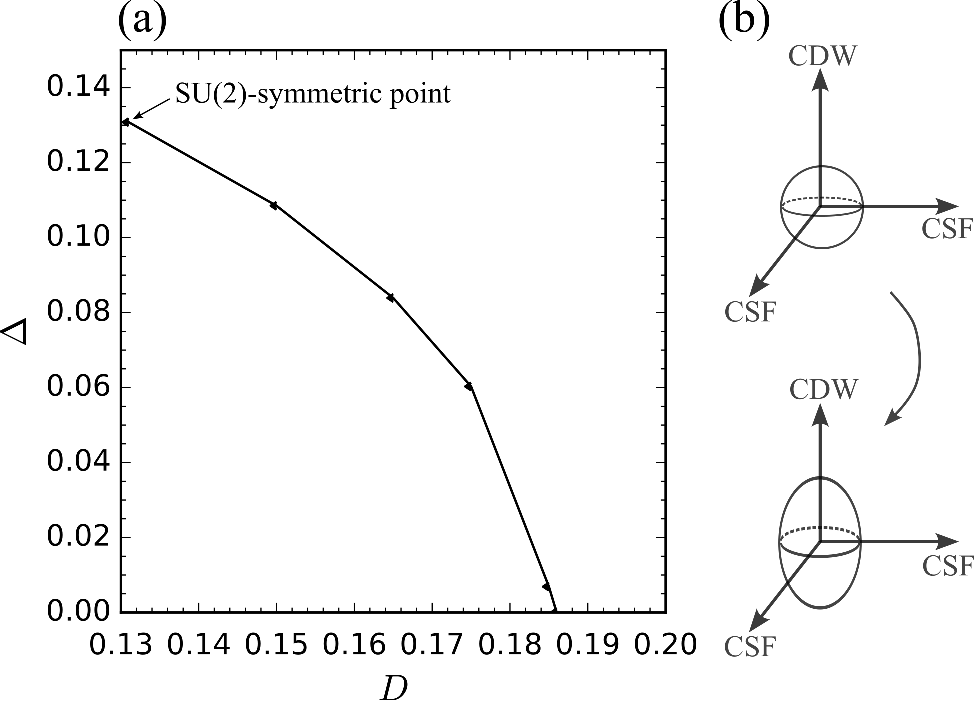}
\caption{(a) The mean-field CSF order $\Delta$ as a fuction of mean-field CDW order $D$ at $T=1/12$ and $|U|=2.0$. (b) Schematic illustration of the lifting of the CDW-CSF degeneracy, resulting in an easy-axis anisotropy.
\label{fig:vertical}
}
\end{figure}

\label{MFCSF}
\textit{Origin of robust supersolidity—} 
To elucidate the mechanism underlying the robust supersolidity in the attractive three-color Hubbard model, we construct a mean-field theory for the reduced system of colors 1 and 2 by tracing out color 3. The resulting mean-field Hamiltonian reads:
\begin{equation}
\label{eq:MFHrealspace}
\begin{aligned}
H^{\mathrm{MF}}_{12}=&-t\sum_{\langle i,j\rangle,\alpha=1,2}(c_{i\alpha}^\dagger c_{j\alpha}+\mathrm{h.c.})-|U|\Delta\sum_i (c_{i2}^{\dagger}c_{i1}^{\dagger}+\mathrm{h.c.})\\
&-|U|D\sum_ie^{i\vec{Q}\cdot\vec{r}_{i}}(n_{i1}+n_{i2}),
\end{aligned}
\end{equation}
where $\Delta=\frac{1}{N}\sum_{i}\langle c_{i2}c_{i1}\rangle$ is the mean-field CSF order parameter, $D=\frac{1}{2N}\sum_{i}\langle n_{i1}+n_{i2}\rangle$ is the mean-field CDW order parameter for colors 1 and 2. Fourier transforming Eq.~(\ref{eq:MFHrealspace}) yields
$H^{\mathrm{MF}}_{12}=\frac{1}{2}\sum_{\vec{k}}\vec{\psi}_{\vec{k}}^\dagger H_{\vec{k}}\vec{\psi}_{\vec{k}}$, with
\begin{equation}
\begin{aligned}
H_{\vec{k}}=\begin{pmatrix}
\epsilon_{\vec{k}} & -|U|D & -|U|\Delta & 0 \\
-|U|D & -\epsilon_{\vec{k}} & 0 & -|U|\Delta \\
-|U|\Delta & 0 & -\epsilon_{\vec{k}} & -|U|D \\
0 & -|U|\Delta & -|U|D & \epsilon_{\vec{k}}
\end{pmatrix},
\end{aligned}
\label{eqHK}
\end{equation}
where $\vec{\psi}_k=(c_{\vec{k},1},c_{\vec{k}+\vec{Q},1},c_{-\vec{k},2}^\dagger,c_{-(\vec{k}+\vec{Q}),2}^\dagger)^T$ is the Nambu spinor and
$\epsilon_{\vec{k}}=-2t[\cos(k_x)+\cos(k_y)]$ is the bare band dispersion. Crucially, 
$D$ couples the $\vec{k}$ and $\vec{k}+\vec{Q}$ momentum sectors, establishing the microscopic basis for the coexistence of the CDW and CSF orders.
\begin{figure}[t]
\includegraphics[width=1.0\linewidth]{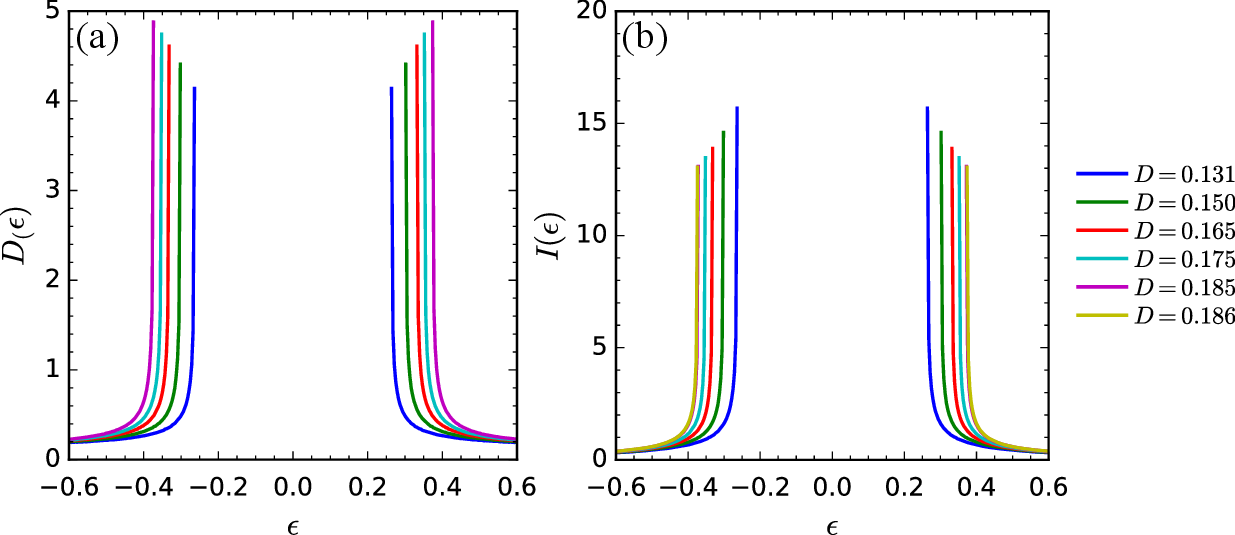}
\caption{(a) Normalized density of states $D(\epsilon)$ and (b) energy-resolved pairing kernel $I(\epsilon)$ as a function of $\epsilon$ at $T=1/12$, $|U|=2.0$ and different values of mean-field CDW order $D$.
\label{fig:DOSSUS}
}
\end{figure}

Diagonalizing $H_k$ in Eq.~(\ref{eqHK}) via a Bogoliubov transformation yields the quasiparticle spectrum $E_{k}=\sqrt{\epsilon_{k}^2+(|U|\Delta)^2+ (|U|D)^2}$.
The self-consistent equation for $\Delta$ reads
\begin{equation}
\label{eq:selfconsistCSF}
\begin{aligned}
\Delta&=\frac{1}{N}\sum_{\vec{k}}\frac{|U|\Delta}{2E_{\vec{k}}}[1-2f(E_{\vec{k}})]\\
&=\frac{1}{N}\int_{\epsilon_{min}}^{\epsilon_{max}}d\epsilon D(\epsilon)\frac{|U|\Delta}{2E_{\vec{k}}}[1-2f(E_{\vec{k}})],     
\end{aligned}
\end{equation}
where $f(E_{\vec{k}})$ is the Fermi-Dirac distribution and $D(\epsilon)$ is the density of states associated with the CDW-renormalized dispersion $\epsilon=\pm\sqrt{\epsilon_k^2+(|U|D)^2}$. Starting from the SU(2)-symmetric point $D=\Delta$, we incrementally increase $D$ and compute the corresponding $\Delta$ at $T=1/12$ and $|U|=2.0$, as shown in Fig.~\ref{fig:vertical} (a).
As $D$ increases, the mean-field CSF order parameter $\Delta$ decreases monotonically, exhibiting a fast lifting of the $\mathrm{SU}(2)$ pseudospin degeneracy in favor of the CDW channel. In the pseudospin picture, this corresponds to the emergence of an easy-axis anisotropy favoring CDW over CSF,  as illustrated in Fig.~\ref{fig:vertical} (b). Consequently, CDW and CSF act as distinct, independent order parameters rather than different projections of a single pseudospin vector. Despite the preference for CDW order, $\Delta$ remains finite over a broad range of $D$, in qualitative agreement with the DQMC results shown in Fig.~\ref{fig:horizontal} (d). 

To elucidate why the CSF order remains robust in the presence of enhanced CDW order, we analyze how the density of states $D(\epsilon)$ and the energy-resolved pairing kernel $I(\epsilon)$ evolve with the mean-field CDW order $D$. Here, $I(\epsilon)$ is defined as
\begin{equation}
I(\epsilon)=\lim_{\Delta\rightarrow 0^{+}}\frac{D(\epsilon)}{2E_{\vec{k}}},
\end{equation}
which quantifies the contribution of states at energy $\epsilon$ to the CSF pairing tendency. Fig.~\ref{fig:DOSSUS} shows $D(\epsilon)$ and $I(\epsilon)$ as a function of $\epsilon$ at $T=1/12$ and $|U|=2.0$ for various values of $D$. As expected, increasing $D$ widens the quasiparticle gap. However, the density of states remains strongly peaked near the gap edges. Consequently, the pairing kernel $I(\epsilon)$ is only moderately suppressed: despite the expanding gap, the large density of states near the gap edges continues to supply a substantial pairing contribution. This explains why CSF order persists over a broad range of CDW order strength on the square lattice at half filling.  

\begin{figure}[t]
\includegraphics[width=0.8\linewidth]{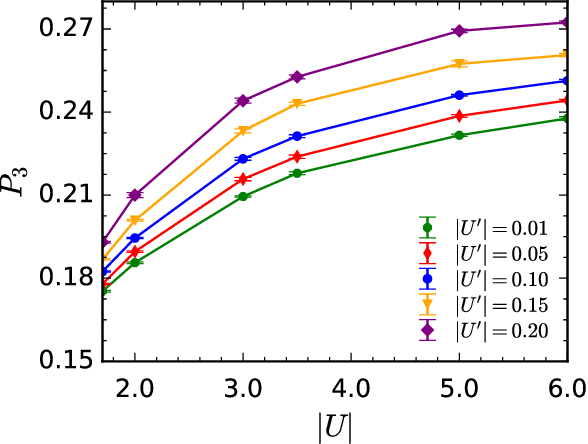}
\caption{The triple occupancy $P_3$ as a function of $|U|$ at different values of $|U'|$ at $T=1/12$. The lattice size is $L=12$.
\label{fig:experiment}
}
\end{figure}

\label{realization}
\textit{Experimental feasibility—} In three-color Fermi gases, three-body loss poses the primary constraint on the stability of many-body states~\cite{LiOttenstein2008,LiHuckans2009,LiSchunck2005}. The on-site triple occupancy, 
\begin{equation}
\label{eq:tripleoccupancy}
P_{3}=\frac{1}{N}\sum_{i}\langle n_{i1}n_{i2}n_{i3}\rangle,
\end{equation}
quantifies local three-body processes and serves as a direct proxy for three-body loss. Fig.~\ref{fig:experiment} displays $P_{3}$ for various values of $|U|$ and $|U'|$. Along the optimal supersolid line ($|U|=2.0$), $P_3$ remains below $0.20$ even up to $|U'|=0.15$. These values fall well within the range of triple occupancies achieved in recent three-color Fermi-Hubbard experiments~\cite{3compMongkolkiattichai2025}, confirming that the supersolid regime identified here resides within an experimentally accessible stability window.

Furthermore, the interaction parameters in our DQMC simulations—characterized by one strong attractive interaction $|U|$ and two weaker attractive interactions $|U'|$—naturally correspond to the low-magnetic-field regime (below $100~\mathrm{G}$) for the relevant scattering lengths in $^6\mathrm{Li}$~\cite{LiHuckans2009}. In this regime, the atom retention fraction after a hold time exceeding $200~\mathrm{ms}$ approaches unity, confirming negligible three-body loss and high state stability~\cite{LiOttenstein2008,LiHuckans2009}.

\label{sec:conclusion}

\textit{Conclusion—} In summary, we establish a viable route to realizing fermionic lattice supersolidity in the half-filled attractive three-color Fermi--Hubbard model on a square optical lattice. This approach is directly compatible with the recent experimental realization of the model~\cite{3compMongkolkiattichai2025}, operating within an interaction regime characterized by low three-body loss, thereby making fermionic supersolidity experimentally accessible on current ultracold-atom platforms.

Mechanistically, coupling to color-3 fermions lifts the $\mathrm{SU}(2)$ pseudospin degeneracy between the CSF and CDW orders, rendering them distinct, competing orders. Most notably, the CSF order remains remarkably robust—sustained by a strong pairing tendency at half filling and sizable phase stiffness at intermediate $|U|$—preventing its rapid suppression. These effects jointly give rise to a broad coexistence region of the CSF and CDW orders, which persists even into the regime where the density order transitions to a trion CDW.

The predicted supersolid state can be experimentally verified through complementary probes of the CDW structure factor~\cite{MMShao2024}, pair coherence~\cite{MMChin2006}, and characteristic energy-gap behavior~\cite{MMLi2024}. Our findings highlight the attractive three-color Fermi--Hubbard model on a simple square lattice as a realistic platform for achieving fermionic lattice supersolidity and probing the interplay of competing density and pairing orders in multicomponent quantum matter.

\acknowledgments
This work is financially supported by the National Natural Science Foundation of China under Grant No. 12574298. We acknowledge the support of the Supercomputing Center of 
Wuhan University.


%

\clearpage
\onecolumngrid
\begin{center}
{\large\bfseries
Supplemental Material: Fermionic Lattice Supersolidity in the Attractive
Three-Color Fermi--Hubbard Model}
\end{center}

\setcounter{figure}{0}
\setcounter{table}{0}
\setcounter{equation}{0}

\renewcommand{\thefigure}{S\arabic{figure}}
\renewcommand{\thetable}{S\arabic{table}}
\renewcommand{\theequation}{S\arabic{equation}}

In this Supplemental Material, we provide the numerical analyses used to establish the phase diagram and the interaction dependence of order parameters presented in the main text, together with an analysis of the occupation number.

\section*{the CSF correlation ratio}
\begin{figure}[thb]
\includegraphics[width=1.0\linewidth]{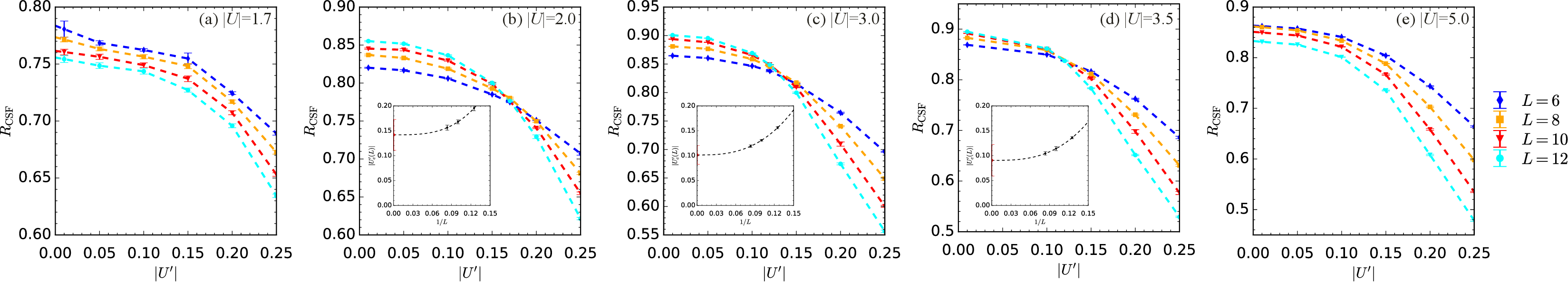}
\caption{
The CSF correlation ratio $R_{\mathrm{CSF}}$ as a function of $|U'|$ for lattice sizes $L=6,8,10,12$ at (a) $|U|=1.7$, (b) $|U|=2.0$, (c) $|U|=3.0$, (d) $|U|=3.5$, and (e) $|U|=5.0$. Insets in (b)-(d) show finite-size extrapolations of the crossing points, with error bars obtained from resampling. The temperature is $T=1/12$.
\label{fig:CSFratio}
}
\end{figure}

To accurately determine the CSF transition at $T=1/12$, we analyze the finite-size behavior of the CSF correlation ratio  $R_{\mathrm{CSF}}$ for different interaction strengths. In Fig.~\ref{fig:CSFratio}, we show $R_{\mathrm{CSF}}$ as a function of $|U'|$ for different system sizes $L$ and interaction strengths $|U|$. 
For $|U|=2.0$–3.5, the curves for different $L$ exhibit clear crossings, indicating a finite critical value of $|U'_c|$. 
We estimate $|U'_c|$ by performing finite-size extrapolations~\cite{3compXu2023} of the crossing points using the form 
\begin{equation}
|U'_c(L)| = |U'_c| + a L^{-b}.
\end{equation}
The extrapolated critical interaction strengths are $|U'_c|$=0.14(3), 0.10(2), and 0.09(3) for $|U|$=2.0, 3.0, and 3.5, respectively.
In contrast, for $|U|$=1.7 and 5.0, no well-defined crossing is observed within the accessible parameter regime, indicating the absence of a finite critical $|U'_c|$.
These results indicate that the region with finite CSF order is confined to an intermediate range of $|U|$, gradually disappearing toward both smaller and larger $|U|$.

\section*{Finite-size extrapolation}
\begin{figure}[t]
\includegraphics[width=0.65\linewidth]{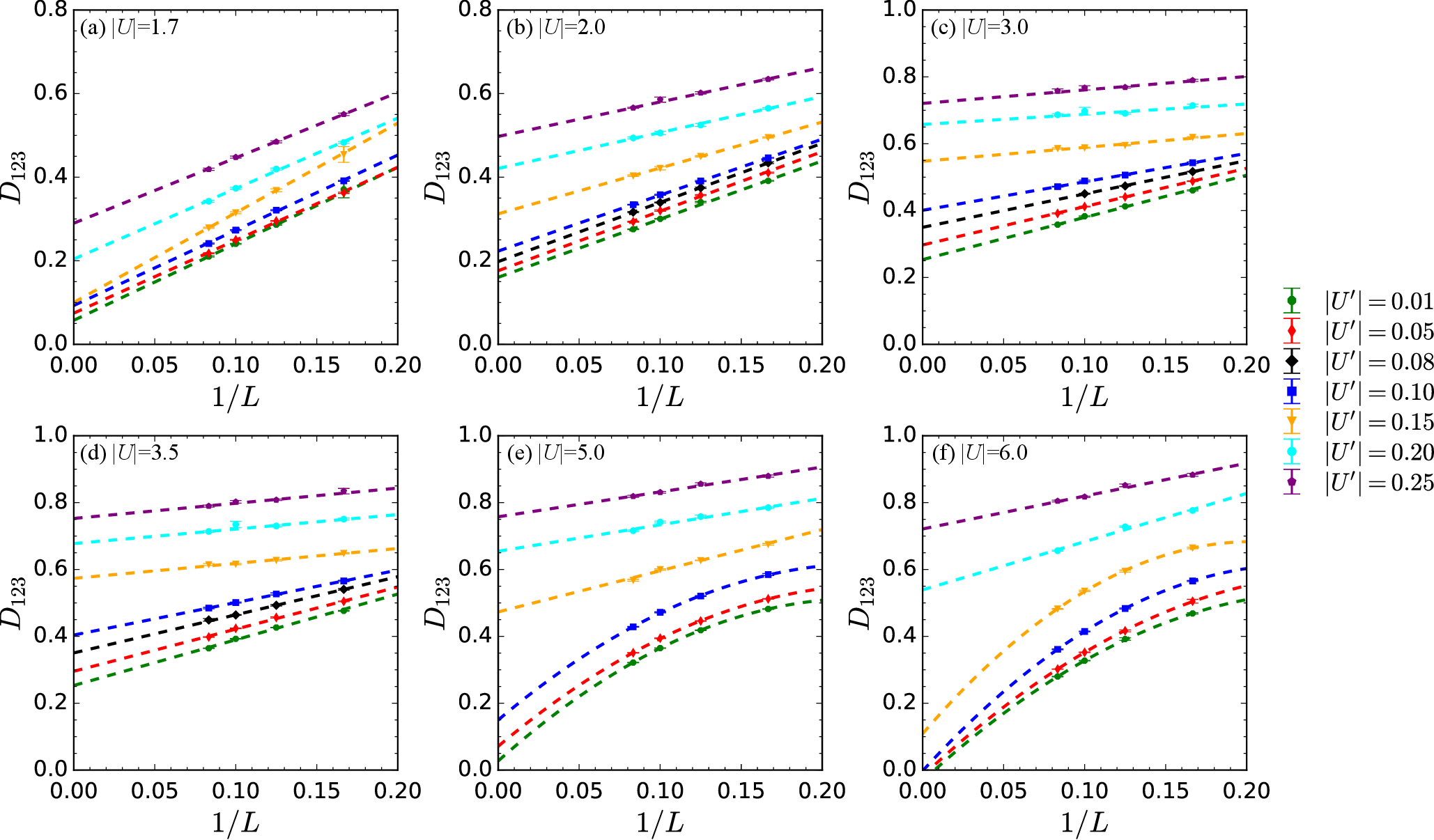}
\caption{
Finite-size extrapolations of the CDW order $D_{123}$ as a function of $1/L$ at (a) $|U|=1.7$, (b) $|U|=2.0$, (c) $|U|=3.0$, (d) $|U|=3.5$, (e) $|U|=5.0$, and (f) $|U|=6.0$. Different symbols and colors represent the data at different values of $|U'|$. The temperature is $T=1/12$. 
\label{fig:CDW123}
}
\end{figure}
\begin{figure}[t]
\includegraphics[width=0.65\linewidth]{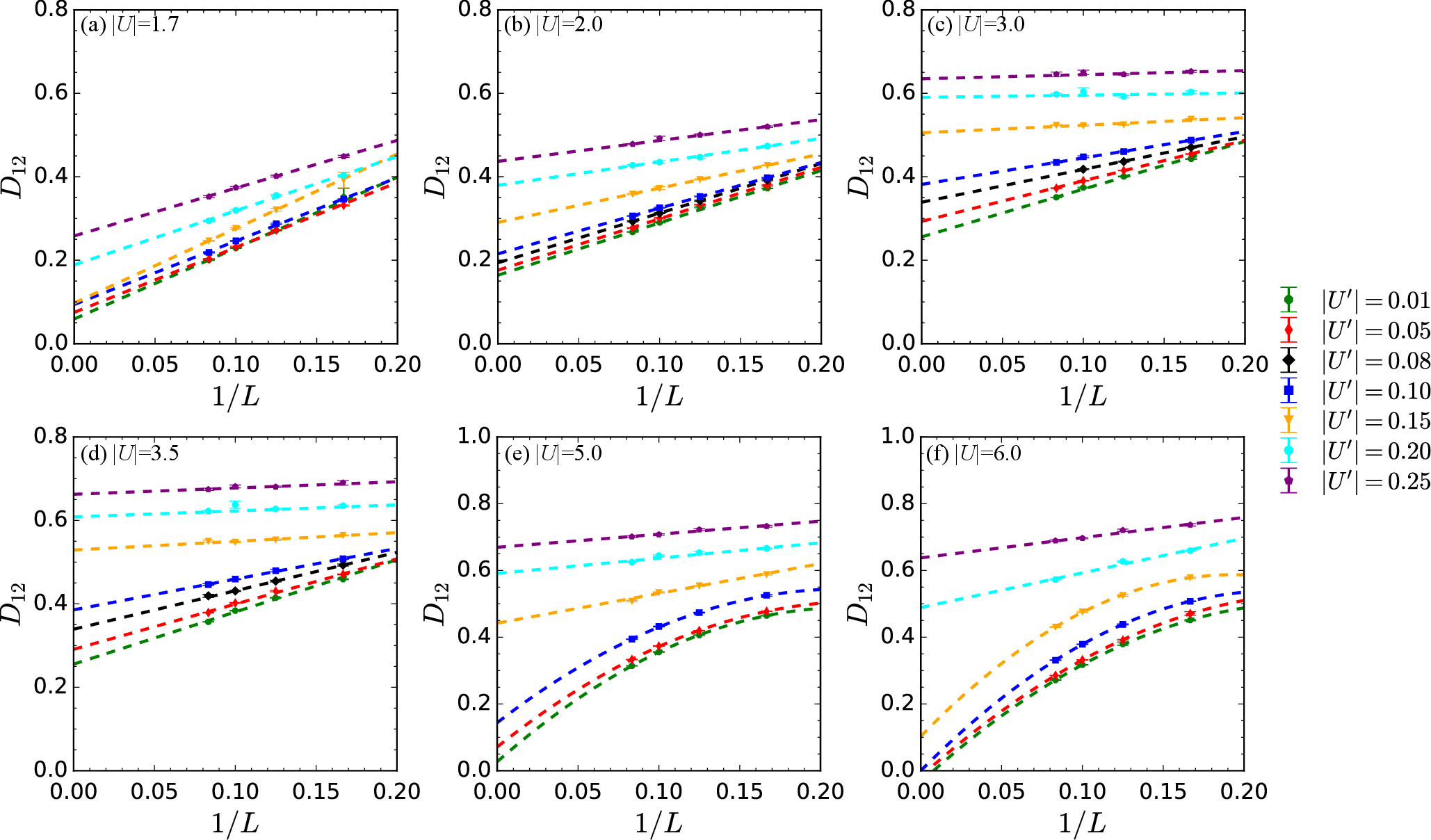}
\caption{
Finite-size extrapolations of the subsystem resolved CDW order $D_{12}$ as a function of $1/L$ at (a) $|U|=1.7$, (b) $|U|=2.0$, (c) $|U|=3.0$, (d) $|U|=3.5$, (e) $|U|=5.0$, and (f) $|U|=6.0$. Different symbols and colors represent the data at different values of $|U'|$. The temperature is $T=1/12$.
\label{fig:CDW12}
}
\end{figure}
Since DQMC simulations are performed on finite lattices, a finite-size extrapolation is required to obtain the thermodynamic-limit values of different order parameters. To obtain their smooth evolution as functions of $|U|$ and $|U'|$, the CDW order $D_{123}$ is extrapolated using
\begin{equation}
\label{extrapolate}
D_{123}=A+\frac{B}{L}+\frac{C}{L^2},
\end{equation}
while other orders are extrapolated using the same form. The extrapolation form Eq.~(\ref{extrapolate}) has been widely adopted in previous quantum Monte Carlo studies of correlated lattice models, including Refs.~\cite{3compXu2023,3compLi2023,3compLi2024,3compLi2025,3compStepp2025,Sandvik1997FiniteSize}.
In practice, however, the dominant finite-size correction is not universal and depends on the finite-size behavior of the numerical data~\cite{Sandvik1997FiniteSize}. Therefore, rather than enforcing a single fitting function throughout the phase diagram, we identify the leading correction directly from the numerical data and employ either linear or quadratic extrapolation accordingly. 
We perform the finite-size extrapolations of $D_{123}$ at different values of $|U|$ and $|U'|$, which are presented in Fig.~\ref{fig:CDW123}. 
The intercept at $1/L \to 0$ gives the thermodynamic-limit value of $D_{123}$ for each $(|U|,|U'|)$. 
For $|U|=1.7-3.5$, the $D_{123}-1/L$ dependence is nearly linear over the entire $|U'|$ range, indicating that the linear correction provides the dominant contribution.
At stronger interactions ($|U|$ = 5.0 and 6.0), noticeable curvature emerges at small $|U'|$, indicating an increasingly important quadratic correction. As $|U'|$ increases, the curvature gradually weakens and the finite-size dependence becomes approximately linear again.

To analyze the crossover from 12-CDW to 123-CDW in the following section, we introduce a subsystem-resolved CDW order parameter,
\begin{equation}
\label{CDW12}
D_{12}=\frac{1}{N}\sqrt{\sum_{ij}e^{i\vec{Q}\cdot (\vec{r}_{i}-\vec{r}_{j})}\sum_{\alpha\beta\in(1,2)}\langle n_{i\alpha}n_{j\beta}\rangle},
\end{equation}
which probes the contribution to the CDW order arising solely from the color-$(1,2)$ subsystem.
Fig.~\ref{fig:CDW12} presents the finite-size extrapolations of $D_{12}$ at different values of $|U|$ and $|U'|$. 
Similar to the case of $D_{123}$, the $D_{12}-1/L$ dependence is nearly linear for $|U|=1.7-3.5$, while noticeable curvature develops at strong $|U|$ and weak $|U'|$. 
Therefore, the same criterion is employed to determine whether linear or quadratic extrapolation is used for the case of $D_{12}$. 

\begin{figure}[t]
\includegraphics[width=0.9\linewidth]{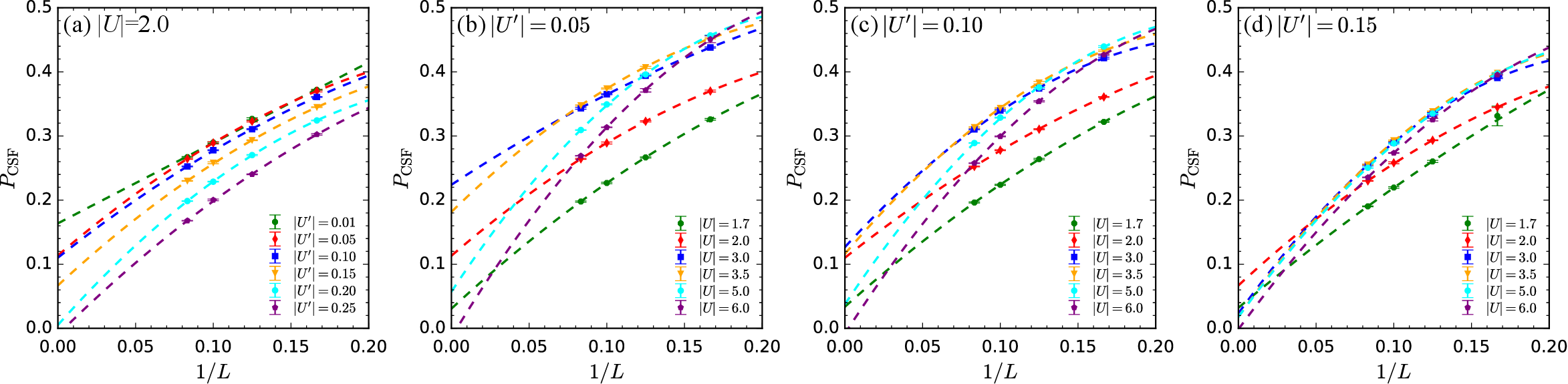}
\caption{ 
Finite-size extrapolations of the CSF order $P_{\mathrm{CSF}}$ as a function of $1/L$ at different values of $|U|$ and $|U'|$. In panel (a), different symbols and colors denote the data at different values of $|U'|$ and fixed $|U|=2.0$, whereas in panels (b)–(d), they denote the data at different values of $|U|$ and fixed $|U'|$. The temperature is $T=1/12$. 
\label{fig:CSFscaling}
}
\end{figure}

Fig.~\ref{fig:CSFscaling} presents the finite-size extrapolations of the CSF order $P_{\mathrm{CSF}}$.
As for the case of CDW orders, the leading finite-size correction is determined directly from the numerical data, and either linear or quadratic extrapolation is adopted accordingly. 
In practice, quadratic extrapolation is used for all datasets except the $|U|$=2.0, $|U'|$ =0.01 case, where the finite-size dependence is sufficiently described by a linear extrapolation.

The extrapolated thermodynamic-limit values are used throughout the main text wherever order parameters enter quantitative analysis. 
Specifically, the extrapolated $D_{123}$ values are interpolated to generate the background color map of the phase diagram. 
The liquid-like phase is then identified as the region where the interpolated thermodynamic-limit values satisfy $D_{123}<0.03$.
Such a small value corresponds to very weak density modulation, consistent with the liquid-like character of this phase. 
In addition, the extrapolated $D_{123}$ and $P_{\mathrm{CSF}}$ values are used to analyze the evolution of orders along the three typical horizontal cuts ($|U'|$ =0.05,0.10,0.15) and the typical vertical cut ($|U|$ =2.0) presented in the main text.

\begin{figure}[b]
\includegraphics[width=0.65\linewidth]{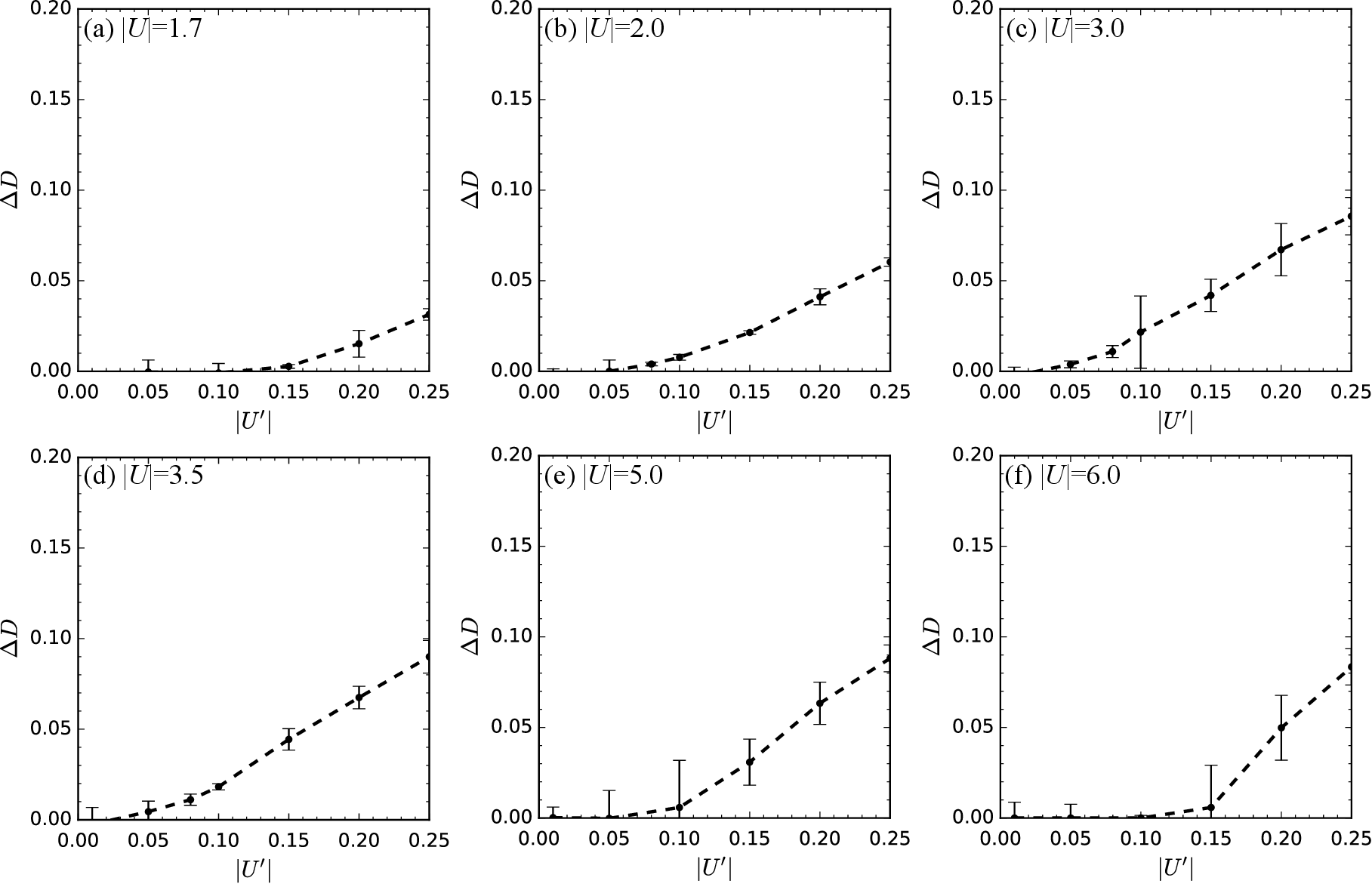}
\caption{
$\Delta D$ as a function of $|U'|$ at (a) $|U|=1.7$, (b) $|U|=2.0$, (c) $|U|=3.0$, (d) $|U|=3.5$, (e) $|U|=5.0$, and (f) $|U|=6.0$. The error bar is $\max[\sigma(D_{123}),\sigma(D_{12})]$. 
\label{fig:CDWdiff}
}
\end{figure}
\section*{the crossover from 12-CDW to 123-CDW}
We characterize the crossover from 12-CDW to 123-CDW by comparing $D_{12}$ and $D_{123}$. For this purpose, we define their difference as
\begin{equation}
\label{DeltaD}
\Delta D=D_{123}-D_{12}.
\end{equation}
When $\Delta D$ is within the uncertainty of $D_{12}$ and $D_{123}$, the global CDW order is dominantly contributed by the 12-CDW; when $\Delta D$ is larger than the uncertainty of $D_{12}$ and $D_{123}$, the contribution of trions becomes non-negligible, signaling the crossover toward 123-CDW. 
Hence, we identify the onset of the crossover from 12-CDW to 123-CDW when
\begin{equation}
\Delta D>\max[\sigma(D_{123}),\sigma(D_{12})].
\end{equation}
In Fig.~\ref{fig:CDWdiff}, we calculate $\Delta D$ at different values of $|U|$ and $|U'|$, with the error bars being
$\max[\sigma(D_{123}),\sigma(D_{12})]$.
As $|U|$ increases from 1.7 to 6.0, the crossover occurs at $|U'|=0.15,0.08,0.05,0.05,0.15,0.20$, respectively.
At $|U|=3.5$, the $|U'|=0.05$ point deserves special discussion. Although at this point, $\Delta D$ is slightly smaller than $\max[\sigma(D_{123}),\sigma(D_{12})]$, it is already clearly positive, and marks the onset of the subsequent monotonic increase of $\Delta D$ with increasing $|U'|$, and is therefore identified as the onset of the crossover at $|U|=3.5$.


\section*{Occupation number}
\begin{figure}[thb]
\includegraphics[width=0.4\linewidth]{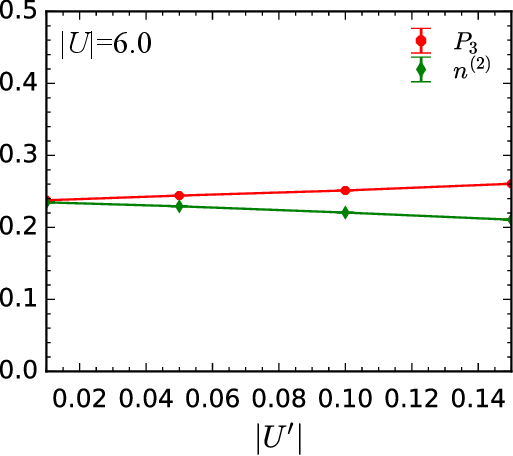}
\caption{
The double-occupation fraction $n^{(2)}$ and triple-occupation fraction $P_{3}$ as a fuction of $|U'|$ at $|U|=6.0$. The temperature is $T=1/12$. 
\label{fig:occupancy}
}
\end{figure}

To characterize the local configurations in the liquid-like (LL) phase, we compare the double-occupation fraction of colors 1 and 2~\cite{3compStepp2025}
\begin{equation}
n^{(2)}=\frac{1}{N}\sum_{i}\langle n_{i1}n_{i2}(1-n_{i3})\rangle,
\end{equation}
with the triple occupancy (triple-occupation fraction) $P_3$ defined in the main text at different values of $|U'|$ and fixed $|U|=6.0$, as shown in Fig.~\ref{fig:occupancy}. In the $|U'|\rightarrow 0$ limit, $n^{(2)}$ and $P_3$ are nearly identical. This reflects the decoupling between the color-(1,2) subsystem and color-3 fermions: particles in the color-(1,2) subsystem are nearly equally likely to occupy a site with or without a color-3 fermion. As $|U'|$ increases, $P_3$ slightly exceeds $n^{(2)}$, indicating that a 12-pair starts to trap a color-3 fermion, forming a trion. Consequently, in the LL phase, there exist several configurations at finite $|U'|$: fermions, doublons, and a small fraction of trions.
\end{document}